\documentclass{iaus}
\usepackage{graphicx}
\usepackage{natbib}

\title[Magnetic Helicity and MHD Dynamics] %% give here short title %%
{Influence of Magnetic Helicity in MHD}

\author[Simon Candelaresi \& Fabio Del Sordo \& Axel Brandenburg]   %% give here short author list %%
{Simon Candelaresi
 \and Fabio Del Sordo
 \and Axel Brandenburg}

\affiliation{NORDITA, AlbaNova University Center,
Roslagstullsbacken 23, SE-10691 Stockholm, Sweden and \\
Department of Astronomy, Stockholm University,
SE 10691 Stockholm, Sweden}

\pubyear{2010}
\volume{271}  %% insert here IAU Symposium No.
\pagerange{369--370}
\jname{Astrophysical Dynamics: From Stars to Galaxies}
\editors{N.H. Brummell, A.S. Brun, M.S. Miesch \& Y. Ponty, eds.}
\doi{10.1017/S1743921311017832}
\begin{document}

\maketitle

\begin{abstract}
Observations have shown that the Sun's magnetic field has helical structures.
The helicity content in magnetic field configurations is a crucial constraint
on the dynamical evolution of the system. Since helicity is connected with
the number of links we investigate configurations
with interlocked magnetic flux rings and one with unlinked rings. It turns out that
it is not the linking of the tubes which affects the magnetic field decay,
but the content of magnetic helicity.
\end{abstract}

% \firstsection % if your document starts with a section,
% \section{Introduction}
Magnetograms of the Sun's surface have shown \citep{Pev95}
that the field lines in the active regions are twisted. From observations of the
chromosphere and the corona it was conjectured \citep{Lek96}
that the magnetic field in sunspots gets twisted before it emerges.
Extrapolation of the three-dimensional structure of
magnetic field lines have shown twist in the field \citep{Gib02}.
Since twisting is connected to helicity we can say that the Sun's magnetic field
shows helical patches.

For two flux tubes which are not twisted and do not intersect each other the magnetic
helicity is related to the number of mutual linking via the formula \citep{Mof69}
$$
H = \int_{V}\textbf{A}\cdot\textbf{B} \ {\rm d} V = 2n\phi_{1}\phi_{2},
$$
where $H$ is the magnetic helicity, $\textbf{B}=\nabla\times\textbf{A}$ is the
magnetic field in terms of the vector potential $\textbf{A}$,
$\phi_{1}$ and $\phi_{2}$ are the magnetic fluxes through the tubes and $n$ is the 
linking number. Since $H$ is a conserved quantity in ideal MHD the linking number
is also conserved.

In ideal MHD and in presence of magnetic helicity the spectral magnetic
energy is bounded from below through the realizability condition \citep{Mof69}
$$
M(k) \ge k|H(k)|/2\mu_{0} \quad {\rm with} \quad
\int M(k) \ {\rm d} k = \langle\textbf{B}^{2}\rangle/2\mu_{0}, \quad
\int H(k) \ {\rm d} k = \langle\textbf{A}\cdot\textbf{B}\rangle
%\label{eq: realizability condition}
$$
and the magnetic permeability $\mu_{0}$ and where $\langle.\rangle$ denotes volume integrals.
It is the aim of our work \citep{DSo10} to study the dynamical evolution of a system
with interlocked flux rings with and without helicity.

\begin{figure}[h]
\begin{minipage}[b]{0.5\linewidth}
\centering
\includegraphics[width=0.2\linewidth]{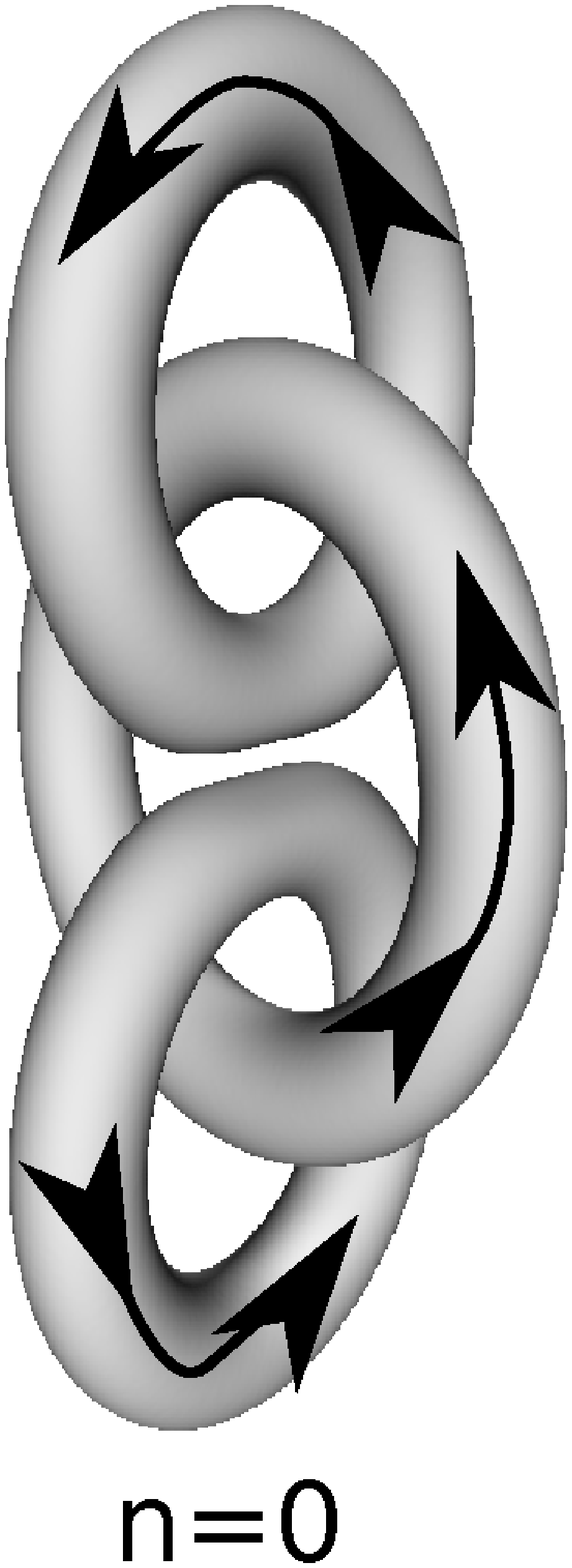} \quad
\includegraphics[width=0.2\linewidth]{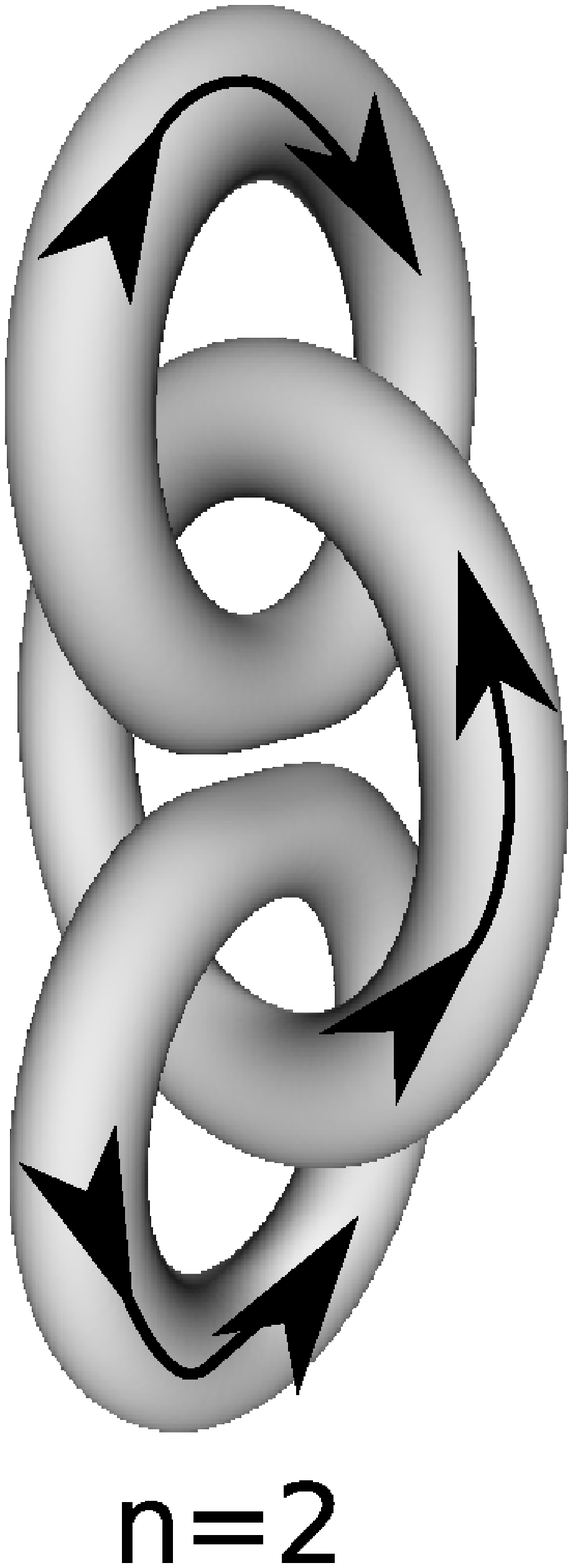} \quad
\includegraphics[width=0.3\linewidth]{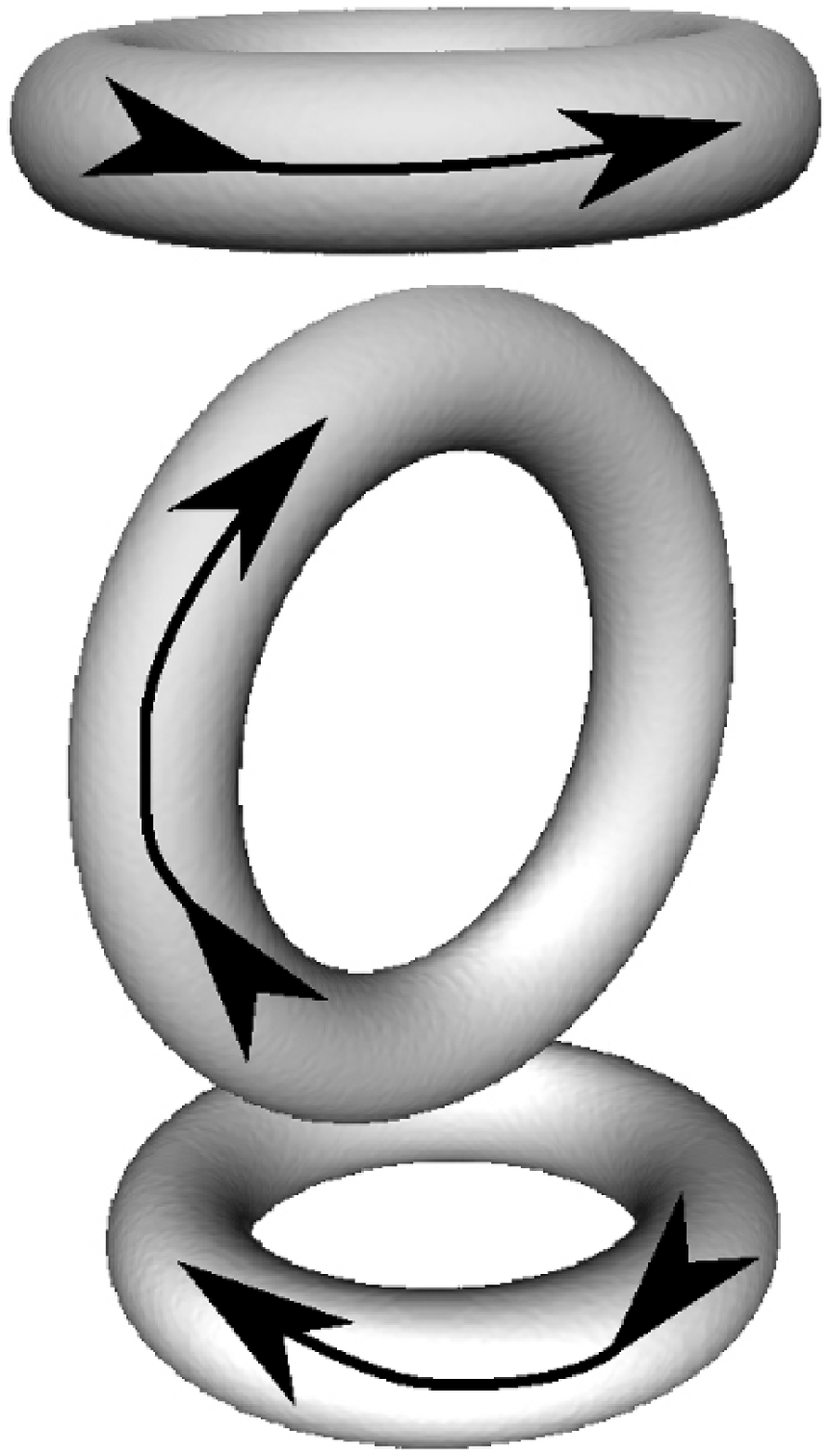}
\caption{The initial triple ring configuration. Interlocked rings with $n=0$
(left) and $n=2$ (center) and non interlocked configuration with
$n=0$ (right). The arrows indicate the direction of the magnetic field.}
\label{fig: initial configuration}
\end{minipage}
\hspace{0.5cm}
\begin{minipage}[b]{0.5\linewidth}
\centering
\includegraphics[width=0.9\linewidth]{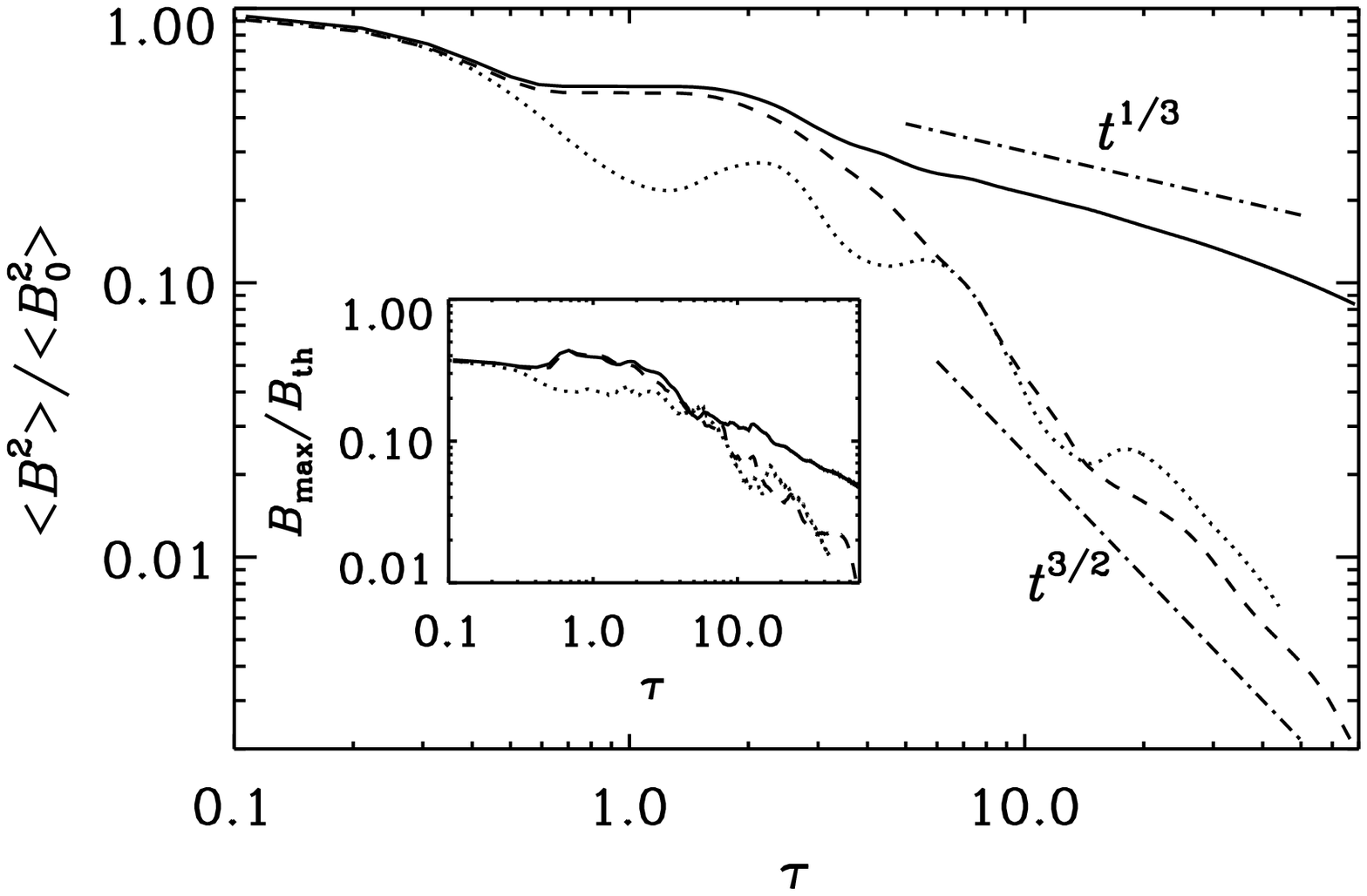} \qquad
\caption{Magnetic energy decay normalized for the initial time
for linking number 2 (solid line), 0 (dashed line) and
the non-interlocked case (dotted line).}
\label{fig: energy decay}
\end{minipage}
\end{figure}

% \section{Model}
The setups under consideration consist of three flux tubes with constant
magnetic flux. For two configurations the rings are interlocked while for
one they are separated. In one of the interlocked configurations we change
the direction of the flux such that the magnetic helicity becomes zero
(Fig.\,\ref{fig: initial configuration}).
The inner ring has a radius which is $1.2$ times larges then for the outer
rings. We solve the full resistive MHD equations for an isothermal and
incompressible medium.
% \begin{eqnarray}
% \frac{\partial\textbf{A}}{\partial t} = \textbf{U}\times\textbf{B} + 
% \eta\nabla^{2}\textbf{A} \\
% \frac{{\rm D}}{{\rm D} t} = -c_{\rm S}^{2} \nabla \ln{\rho} +
% \textbf{J}\times\textbf{B}/\rho + \textbf{F}_{\rm visc} \\
% \frac{{\rm D}\ln{\rho}}{{\rm D}t} = -\nabla\cdot\textbf{U}
% \end{eqnarray}
% where $\eta$ is the magnetic diffusivity, $c_{\rm S}$ the sound
% speed, $\textbf{U}$ the velocity, $\rho$ the density, $\textbf{J}$
% the current and $\textbf{F}_{\rm visc}$ the viscous force.
% 
% The times will be given in Alfv\'en times $T_{\rm A} = \sqrt{\mu_{0}\rho_{0}}
% R_{0}^{3}/\phi$. Where $R_{0}$ is the size of one ring.

% \section{Results}
For the interlocked configuration with finite magnetic helicity
the magnetic energy decays with a
power law of $t^{-1/2}$, while both non-helical configurations decay
much faster with a power law of $t^{-3/2}$
(Fig.\,\ref{fig: energy decay}).

\begin{figure}[h!]
\begin{center}
 \includegraphics[width=0.45\linewidth]{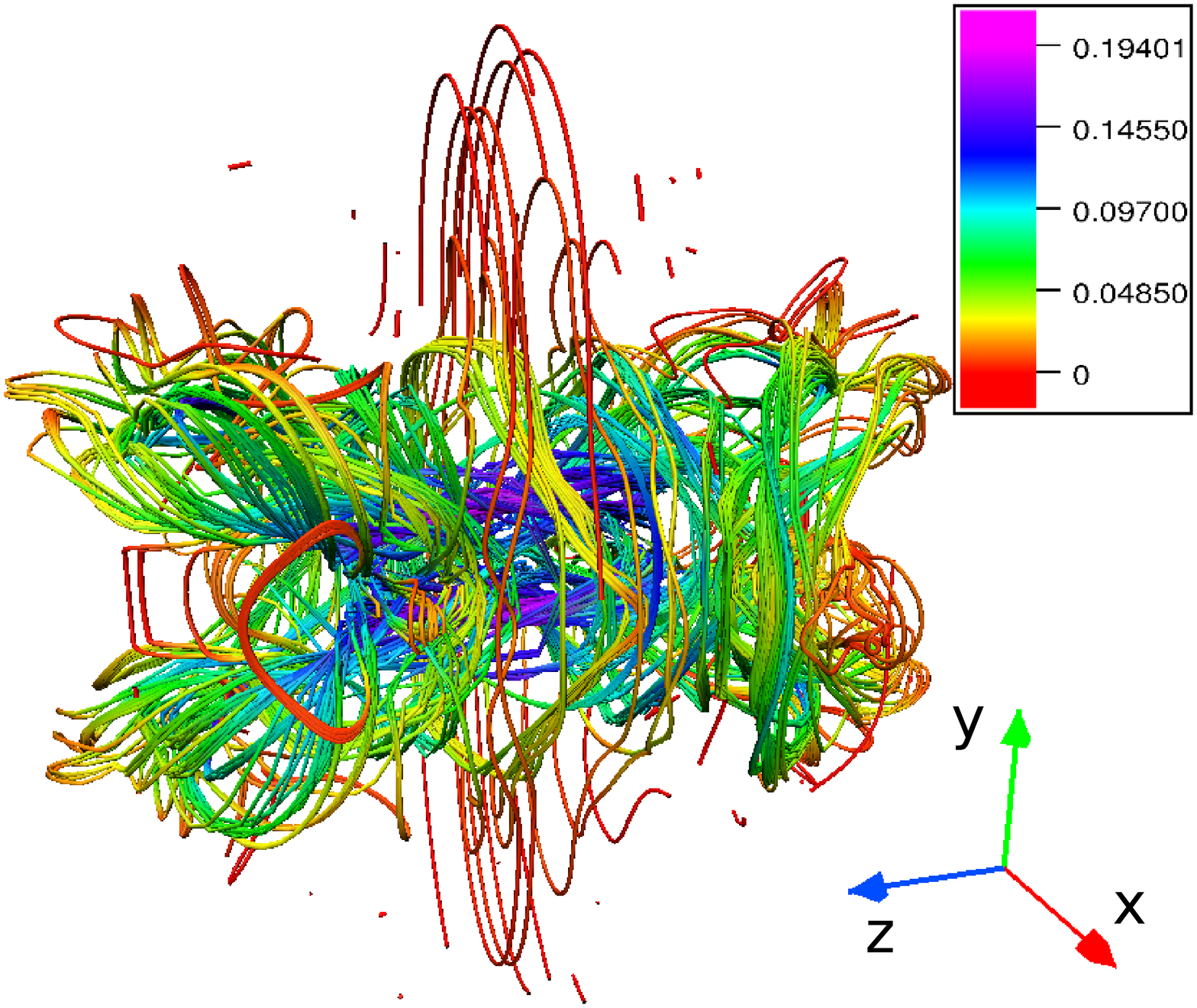} \qquad
 \includegraphics[width=0.45\linewidth]{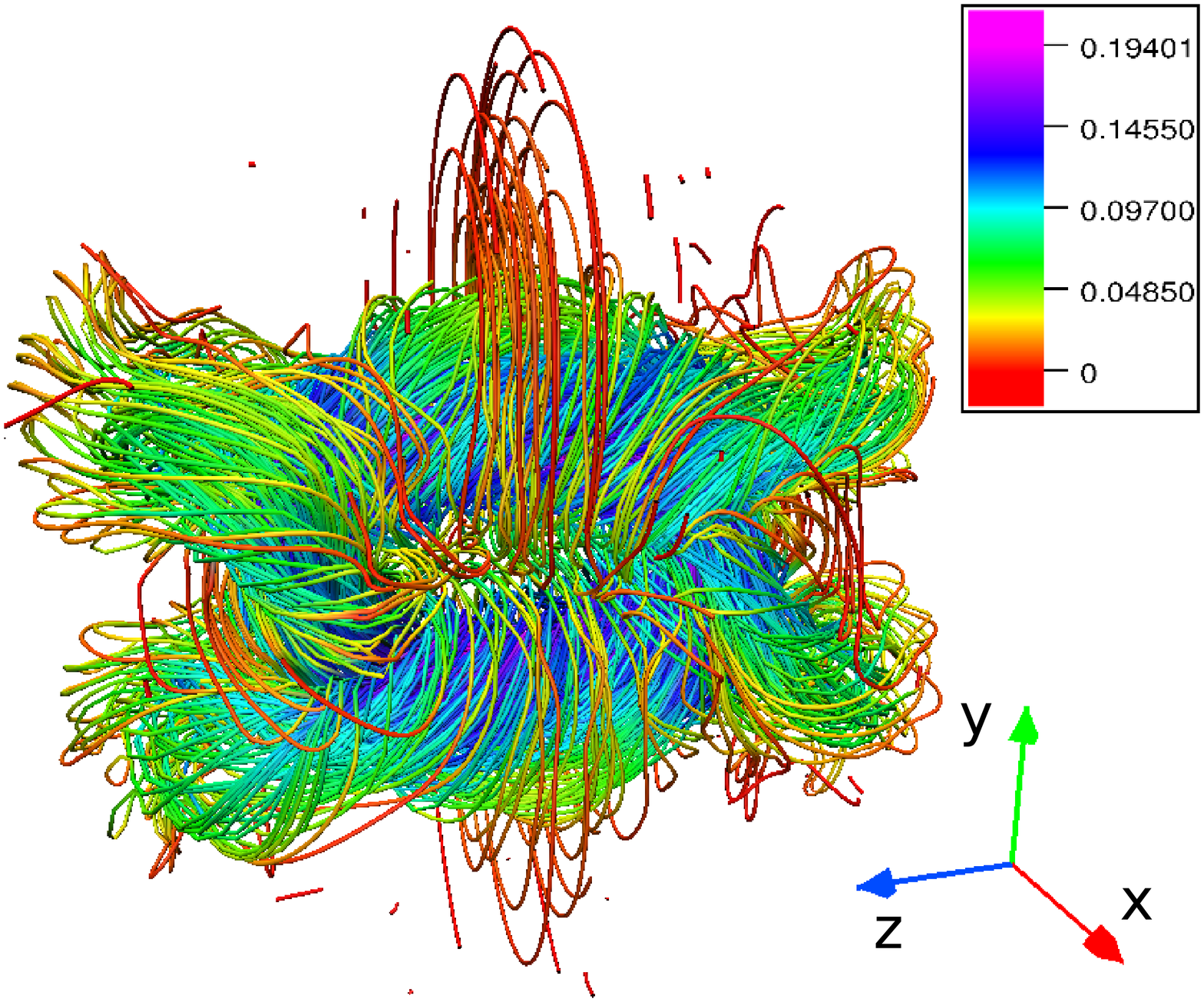}
 \caption{Magnetic field configuration after 4 Alfv\'en times for the case
          of linking number zero (left) and finite linking number (right).
          The colors represent the magnitude of the magnetic field.}
 \label{fig: field lines evolution}
\end{center}
\end{figure}

One can see (Fig.\,\ref{fig: field lines evolution}) that
the linked structure is conserved for the finite helicity case.
For even later times the linking gets transformed into twisting.

% \section{Conclusions}
Due to the realizability condition
the decay of magnetic energy is slowed down by the magnetic helicity which
decays slowly.
Since helicity is decaying slowly and is almost conserved also the number of
linkings is almost conserved. This is why the linking is transformed into
twisting of the fields which then contributes to the helicity.
The test run with non-interlocked rings shows that it is the helicity content
of the system and not the actual number of linkings which affects the decay.

\end{document}